\documentclass[fleqn,usenatbib]{mnras}

\usepackage{newtxtext,newtxmath}

\usepackage[T1]{fontenc}
\usepackage{ae,aecompl}

\usepackage{graphicx}	
\usepackage{amsmath}	

\usepackage{amssymb}	
\usepackage{multicol}
\usepackage{bm}
\usepackage{booktabs}
\usepackage{cleveref}
\usepackage{natbib}

\defcitealias{Bian2018}{B18}
\defcitealias{PP04}{PP04}

\title[The MZR and FMR at $z\sim1.5$]{The MOSDEF Survey: The Mass-Metallicity relationship and the existence of the FMR at $\lowercase{z}\sim1.5$ $^{1}$}

\author[M. W. Topping et al.]{Michael W. Topping,$^{2,3}$\thanks{E-mail: mtopping@astro.ucla.edu}
Alice E. Shapley,$^{2}$
Ryan L. Sanders,$^{4,5}$
Mariska Kriek,$^{6}$ \newauthor
Naveen A. Reddy,$^{7}$
Alison L. Coil,$^{8}$
Bahram Mobasher,$^{7}$
Brian Siana,$^{7}$\newauthor
William R. Freeman,$^{7}$
Irene Shivaei,$^{3,5}$
Mojegan Azadi,$^{9}$
Sedona H. Price,$^{10}$\newauthor
Gene C. K. Leung,$^{8}$
Tara Fetherolf,$^{7}$
Laura de Groot,$^{11}$
Tom Zick,$^{6}$\newauthor
Francesca M. Fornasini,$^{9}$
Guillermo Barro$^{12}$
Jordan N. Runco,$^{2}$
\\
$^{1}$Based on data obtained at the W.M. Keck Observatory, which is operated as a scientific partnership among the California Institute of \\ Technology, the University of California,  and the National Aeronautics and Space Administration, and was made possible by the generous  \\ financial support  of the W.M. Keck Foundation.\\
$^{2}$Physics \& Astronomy Department, University of California: Los Angeles, 430 Portola Plaza, Los Angeles, CA 90095, USA\\
$^{3}$Department of Astronomy / Steward Observatory, University of Arizona, 933 N Cherry Ave, Tucson, AZ 85721\\
$^{4}$Department of Physics, University of California, Davis, One Shields Ave, Davis, CA 95616, USA\\
$^{5}$Hubble Fellow\\
$^{6}$Astronomy Department, University of California, Berkeley, CA 94720, USA\\
$^{7}$Department of Physics \& Astronomy, University of California, Riverside, 900 University Avenue, Riverside, CA 92521, USA\\
$^{8}$Center for Astrophysics and Space Sciences, University of California, San Diego, 9500 Gilman Dr., La Jolla, CA 92093-0424, USA\\
$^{9}$Harvard-Smithsonian Center for Astrophysics, 60 Garden Street, Cambridge, MA, 02138, USA \\
$^{10}$Max-Planck-Institut f\"ur Extraterrestrische Physik, Postfach 1312, Garching, 85741, Germany \\
$^{11}$Department of Physics, The College of Wooster, 1189 Beall Avenue, Wooster, OH 44691, USA\\
$^{12}$Department of Phyics, University of the Pacific, 3601 Pacific Ave, Stockton, CA 95211, USA
}

\begin{document}

\label{firstpage}
\pagerange{\pageref{firstpage}--\pageref{lastpage}}
\maketitle

\begin{abstract}

We analyze the rest-optical emission-line ratios of $z$$\sim$$1.5$ galaxies drawn from the MOSFIRE Deep Evolution Field (MOSDEF) survey. Using composite spectra we investigate the mass-metallicity relation (MZR) at $z$$\sim$$1.5$ and measure its evolution to $z$$=$$0$. When using gas-phase metallicities based on the N2 line ratio, we ﬁnd that the MZR evolution from $z$$\sim$$1.5$ to $z$$=$$0$ depends on stellar mass, evolving by $\Delta\rm log(\rm O/H)$$\sim$$0.25$ dex at $M_* $$<$$10^{9.75}M_{\odot}$ down to $\Delta\rm log(\rm O/H)$$\sim$$0.05$ at $M_*$$ \gtrsim$$ 10^{10.5}M_{\odot}$. In contrast, the O3N2-based MZR shows a constant offset of $\Delta\rm log(\rm O/H)$$\sim$$0.30$ across all masses, consistent with previous MOSDEF results based on independent metallicity indicators, and suggesting that O3N2 provides a more robust metallicity calibration for our $z$$\sim$$1.5$ sample. We investigated the secondary dependence of the MZR on SFR by measuring correlated scatter about the mean $M_*$-speciﬁc SFR and $M_*$$-$$\log(\rm O3N2)$ relations. We ﬁnd an anti-correlation between $\log(\rm O/H)$ and sSFR offsets, indicating the presence of a $M_*$-SFR-Z relation, though with limited signiﬁcance. Additionally, we ﬁnd that our $z$$\sim$$1.5$ stacks lie along the $z$$=$$0$ metallicity sequence at ﬁxed $\mu$$=$$\log(M_*/M_{\odot})$$ -$$ 0.6 $$\times$$ \log(\rm SFR / M_{\odot} yr^{-1})$ suggesting that the $z$$\sim$$1.5$ stacks can be described by the $z$$=$$0$ fundamental metallicity relation (FMR). However, using different calibrations can shift the calculated metallicities off of the local FMR, indicating that appropriate calibrations are essential for understanding metallicity evolution with redshift. Finally, understanding how [NII]/H$\alpha$ scales with galaxy properties is crucial to accurately describe the effects of blended [NII] and H$\alpha$ on redshift and H$\alpha$ ﬂux measurements in future large surveys utilizing low-resolution spectra such as with \textit{Euclid} and the \textit{Roman Space Telescope}.
\end{abstract}

\section{Introduction} 
\label{sec:intro}

Gas-phase metallicity provides a view of the integrated evolution of many key physical processes within galaxies.  These include the production of heavy metals through star formation after which these metals are distributed throughout the galaxy \citep{Matteucci2012, Nomoto2013}.  The metallicity is further regulated by the exchange of material with the galaxy environment, either by the expulsion of enriched material into the surroundings, or by metal-rich or pristine inflows \citep{Tremonti2004, Erb2006, Steidel2010,  Tumlinson2011, Chisholm2018}.  The metallicity in the interstellar medium (ISM) is therefore a crucial window into the current evolutionary state of the galaxy.

Due to the importance of understanding the metallicity within galaxies, many different methods for measuring the gas-phase oxygen abundance have been devised. Ideally, one would use a so-called `direct' method, which relies on observations of faint auroral lines and another strong line of the same species. The electron temperature can then be measured, and, in combination with the electron density, the ionic abundances are calculated based on atomic physics.  However, the faintness of these auroral lines means that this method can primarily only be used in nearby HII regions and galaxies \citep[e.g.,][]{Izotov2006, AM13, Curti2020, Berg2020}.  Another very common method requires the measurement of multiple nebular emission lines which are then formulated into line ratios such as $\rm [NII]\lambda 6584/H\alpha$ and $(\rm [OIII]\lambda 5007/H\beta)/([NII]\lambda 6584/H\alpha)$, hereafter abbreviated by N2 and O3N2, respectively \citep[e.g.,][]{PP04, Marino2013}.  There are several advantages to these strong-line ratios beyond the ease of observing bright lines. In the case of N2 and O3N2, the proximity in wavelength of the compared lines negates the need for accurate dust corrections as there is little differential extinction.  However, this method relies on the calibration of strong-line ratios to oxygen abundance, frequently using the direct method observations of local HII regions and galaxies \citep[e.g.,][]{PP04, Curti2020}, which may not be applicable at high redshift.

The understanding of metallicities for a comprehensive set of galaxies at high redshift has progressed due to the advent of large multiplexed near-IR spectrographs on large telescopes \citep[e.g.,][]{Kriek2015, Steidel2014, Silverman2015}. Measuring the metallicity of an individual galaxy is a useful tool for studying its evolution. However the metallicities of a population of galaxies informs the current evolutionary state of the universe \citep{Tremonti2004, Kewley2008, AM13}.  The mass-metallicity relationship (MZR) has been observed in galaxies at high redshift out to at least $z\sim3.5$, and indicates that metallicity correlates with stellar mass \citep{Erb2006, Maiolino2008, Steidel2014,  Sanders2015, Onodera2016, Sanders2018, Gillman2021}. In the local universe, studies of the MZR found that the scatter in this relationship could be reduced when extending the relation using the star-formation rate (SFR) dimension, resulting in a $\rm M_*$-SFR-Z relation called the ``fundamental metallicity relation''  \citep[FMR;][]{Ellison2008, Mannucci2010, Lara-Lopez2010, Yates2012, Salim2014}. The FMR has been interpreted as the regulation of star formation and gas-phase metallicity content by gas flows such that inflowing pristine material dilutes the ISM and increases star formation \citep[e.g.,][]{Dave2017}.  Analysis of galaxies up to $z\sim3$ have provided evidence for a mass-metallicity-SFR relation similar to that observed in local galaxies.  However, there has been much controversy over whether this FMR remains relatively consistent through cosmic time \citep{Wuyts2016, Maier2014, Cresci2019, Sanders2020} or if it evolves significantly with redshift \citep{Cullen2014,Zahid2014b, Troncoso2014, Kashino2017}.

Recent results suggest that the FMR is constant up to high redshift \citep{Cresci2019, Sanders2020}, and any variation of the MZR observed at different redshifts represents a different cross section of the unchanging FMR associated with the evolution of the SFR-$M_*$ relation with redshift.  Thus, while the MZR observed at high redshift and in the local universe are different as galaxies at high redshift are more actively star-forming and gas rich, galaxies at both epochs exhibit the same \textit{equilibrium} between gas flows and star formation \citep{Mannucci2010, Lilly2013}. However, more data are needed throughout cosmic time in order to establish the invariance of the FMR as a function of redshift. Understanding this evolution, if any, by sampling the FMR across different redshifts, and spanning a broad range of galaxy properties, is key for assembling complete galaxy chemical evolution models. Furthermore, as metallicity calibrations play a crucial role in establishing the FMR, and the choice of calibration can drastically influence the observed relations, care must be taken to use consistent measurements at different redshifts. Lastly, it is important to establish how the evolution of the MZR and FMR varies as a function of stellar mass, as it appears that high-mass galaxies require less evolution in metallicity toward lower redshifts \citep{Zahid2011, Moustakas2011, Zahid2013, PerezMontero2014, Suzuki2017}.

In this paper we investigate the MZR and FMR of galaxies at $z\sim1.5$ from the MOSDEF survey \citep{Kriek2015}.  In Section 2 we describe the data and sample statistics, and provide an overview of our methods.  Section 3 presents the results of our analysis.  Section 4 provides a discussion of our results. Finally, Section 5 provides a summary and some conclusions.  Throughout this paper we assume a cosmology with $\Omega_m = 0.3$, $\Omega_{\Lambda}=0.7$, $H_0=70 \textrm{km s}^{-1}\ \textrm{Mpc}^{-1}$, and adopt solar abundances from \citet[][i.e., $Z_{\odot}=0.014$, $12+\log(\rm O/H)_{\odot}=8.69$]{Asplund2009}.

\begin{figure*}
    \centering
    \includegraphics[width=1.0\linewidth]{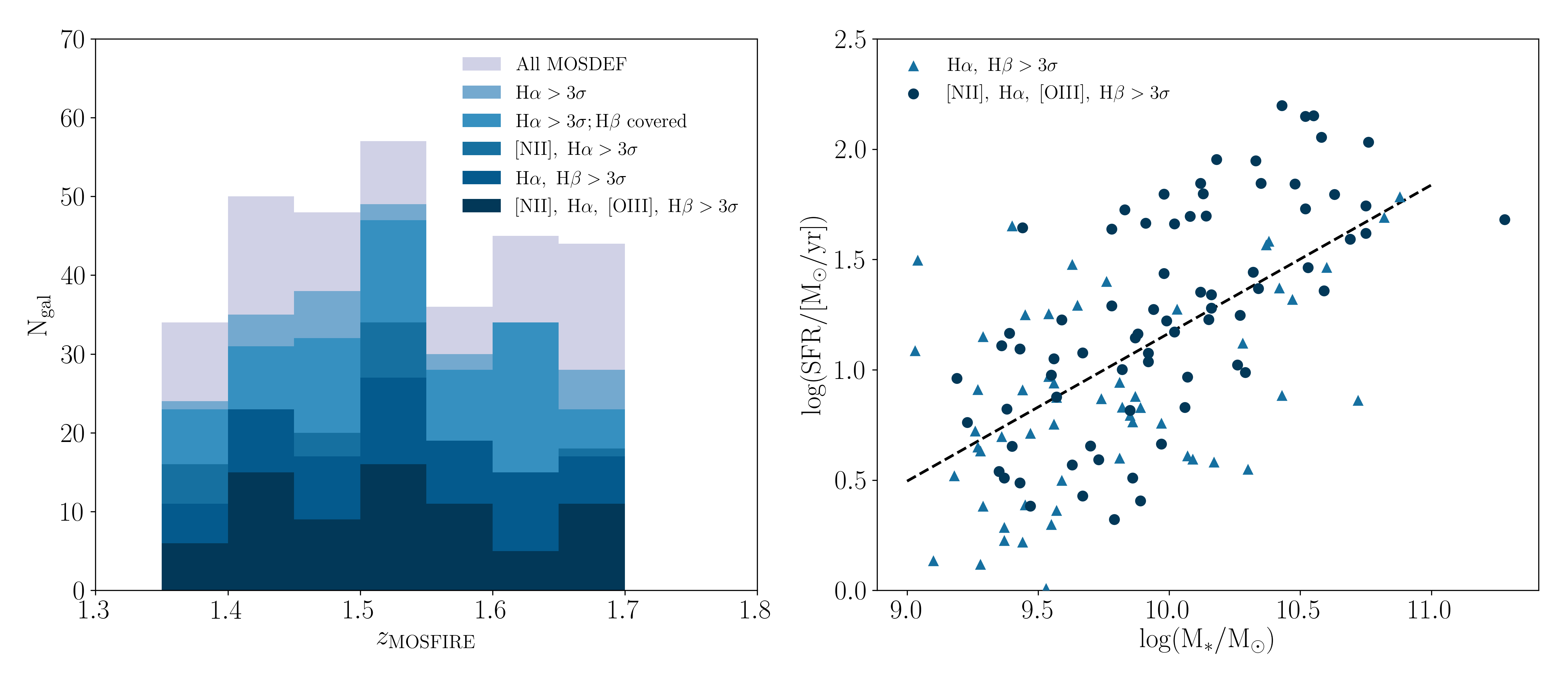}
    \caption{\textit{Left}: Redshift histogram for galaxies in our sample and part of the MOSDEF survey with $1.37 \le z \le 1.7$.  Throughout this paper we define several different samples based on the which strong lines are detected at $\ge 3 \sigma$.  Table~\ref{tab:samplenumbers} lists the size of each subsample shown here. \textit{Right}: The H$\alpha$-based SFR vs. stellar mass for each galaxy in our sample that has a $\ge 3 \sigma$ detection in at least H$\alpha$ and H$\beta$ in order to calculate reliable dust-corrected SFRs.  Galaxies that have $>3\sigma$ detections in H$\alpha$, [NII], H$\beta$, and [OIII] are indicated by circles, and galaxies with $>3\sigma$ detections in H$\alpha$ and H$\beta$, but not [NII] and [OIII], are shown as triangles.  The dashed line shows the best-fit relation calculated for all galaxies with $>3\sigma$ detections in at least H$\alpha$ and H$\beta$.} 
    \label{fig:sample}
\end{figure*}

\section{Data and Measurements}
\label{sec:data}

\subsection{The MOSDEF Survey}

The measurements presented in this paper utilize rest-optical spectra of galaxies at $z\sim1.5$ from the MOSDEF survey \citep{Kriek2015}, collected over a period of 48.5 nights during 2012--2016 using the Multi-Object Spectrometer for Infra-Red Exploration \citep[MOSFIRE;][]{McLean2012}. The full spectroscopic sample of MOSDEF comprises near-infrared spectra of $\sim1500$ $H$-band selected galaxies targeted to lie within three distinct redshift intervals ($1.37 \le z \le 1.70$, $2.09 \le z \le 2.61$, and $2.95 \le z \le 3.80 $).  Based on the scatter between photometric and spectroscopic redshifts of the MOSDEF targets, the actual redshift ranges slightly differ from the initial target ranges, such that the true redshift ranges are $1.37 \le z < 1.90$, $1.90 \le z \le 2.65$, and $2.95 \le z \le 3.80 $. This work makes use of the extensive ancillary datasets available for the MOSDEF targets, including measurements from the CANDELS \citep{Grogin2011} and 3D-HST \citep{Momcheva2016} surveys. The moderate resolution ($\textrm{R}\sim3500$) MOSDEF spectra were analyzed in order to obtain redshift and flux information for all rest-optical emission lines detected and that lie within the $Y$, $J$, $H$, and $K$ bands, the strongest of which are: [OII]$\lambda3727$, H$\beta$, [OIII]$\lambda \lambda 4959,5007$, H$\alpha$, [NII]$\lambda6584$, and [SII]$\lambda \lambda 6717,6731$. However, in the redshift range of galaxies studied in this work ($1.37 \le z \le 1.70$) we primarily use the H$\beta$, [OIII]$\lambda \lambda 4959,5007$, H$\alpha$, and [NII]$\lambda6584$ lines in the $J$ and $H$ bands.

\begin{table*}
\begin{center}
\renewcommand{\arraystretch}{1.4}
\begin{tabular}{llrrr}
\toprule
  Wavelength coverage & Detected lines ($\ge 3\sigma$) & $N_{\rm gal}$ & $\log(M_*/\rm M_{\odot})_{\rm med}$ & $\log(\rm SFR/M_{\odot} yr^{-1})_{\rm med}$  \\
\midrule
 \multicolumn{2}{c}{All MOSDEF ($1.37 < z < 1.7$)}  & 314 & 9.97 & -  \\
 \midrule
 H$\alpha$, [NII]& H$\alpha$ & 238 & 9.98 & - \\
H$\alpha$, [NII]& H$\alpha$, [NII] & 138 & 10.2 & - \\
\midrule
H$\alpha$, [NII], H$\beta$, [OIII]& H$\alpha$ & 218  & 9.88 & - \\
H$\alpha$, [NII], H$\beta$, [OIII]& H$\alpha$, H$\beta$ & 129  & 9.89 & 1.10 \\
H$\alpha$, [NII], H$\beta$, [OIII]& H$\alpha$, [NII], H$\beta$, [OIII] & 73 & 10.02 & 1.27 \\

 \bottomrule
 \end{tabular}
 \end{center}
 \caption{Summary of galaxy properties for the subsamples defined in Section~\ref{sec:samples}.}

\label{tab:samplenumbers}
\end{table*}

\begin{table}
\begin{center}
\renewcommand{\arraystretch}{1.4}
\begin{tabular}{rrr}
\toprule
\multicolumn{3}{c}{$\rm [NII]/H\alpha\ M_* \ stacks$}\\
\midrule
    $N_{\rm gal}$ & $\log(M_*/\rm M_{\odot})_{\rm med}$ $^{\rm a}$ & $\log(\rm [NII]/H\alpha)$ $^{\rm b}$ \\
\midrule
$30$ & $9.26_{- 0.06 }^{ +0.01 }$ & $-1.14_{- 0.10 }^{ +0.11 }$\\
$30$ & $9.44_{- 0.01 }^{ +0.04 }$ & $-1.12_{- 0.10 }^{ +0.15 }$\\
$30$ & $9.64_{- 0.01 }^{ +0.05 }$ & $-1.05_{- 0.05 }^{ +0.13 }$\\
$30$ & $9.82_{- 0.01 }^{ +0.02 }$ & $-0.86_{- 0.07 }^{ +0.07 }$\\
$30$ & $9.98_{- 0.03 }^{ +0.02 }$ & $-0.70_{- 0.08 }^{ +0.02 }$\\
$30$ & $10.23_{- 0.04 }^{ +0.02 }$ & $-0.62_{- 0.05 }^{ +0.04 }$\\
$29$ & $10.48_{- 0.04 }^{ +0.03 }$ & $-0.48_{- 0.05 }^{ +0.01 }$\\
$29$ & $10.78_{- 0.04 }^{ +0.06 }$ & $-0.51_{- 0.03 }^{ +0.04 }$\\

\toprule
\toprule
\multicolumn{3}{c}{$\rm O3N2\ M_* \ stacks$}\\
\midrule
    $N_{\rm gal}$ & $\log(M_*/\rm M_{\odot})_{\rm med}$ $^{\rm a}$ & $\log(\rm O3N2)$ $^{\rm b}$ \\
\midrule

$28$ & $9.20_{- 0.06 }^{ +0.01 }$ & $1.78_{- 0.18 }^{ +0.11 }$\\
$28$ & $9.40_{- 0.04 }^{ +0.00 }$ & $1.64_{- 0.20 }^{ +0.13 }$\\
$27$ & $9.54_{- 0.03 }^{ +0.01 }$ & $1.50_{- 0.13 }^{ +0.22 }$\\
$27$ & $9.68_{- 0.02 }^{ +0.04 }$ & $1.48_{- 0.17 }^{ +0.01 }$\\
$27$ & $9.83_{- 0.01 }^{ +0.02 }$ & $1.28_{- 0.13 }^{ +0.08 }$\\
$27$ & $9.98_{- 0.04 }^{ +0.01 }$ & $1.02_{- 0.04 }^{ +0.13 }$\\
$27$ & $10.18_{- 0.03 }^{ +0.04 }$ & $0.92_{- 0.12 }^{ +0.07 }$\\
$27$ & $10.58_{- 0.06 }^{ +0.03 }$ & $0.62_{- 0.05 }^{ +0.11 }$\\

 \bottomrule
 \end{tabular}
 \end{center}
 \caption{Stellar mass and emission-line ratios of $z\sim1.5$ composite spectra. }
 {$^{\rm a}$}{ Median stellar mass of galaxies in the bin.}\\
 {$^{\rm b}$}{ Line ratio measured from the composite spectrum.}

\label{tab:Mstacks}
\end{table}

\subsection{Galaxy properties and measurements}
In this study we analyze several global galaxy properties.  The procedure for estimating stellar masses for galaxies in our sample is described in detail in \citet{Kriek2015}.  In brief, stellar masses were derived using the fitting code \textsc{fast} \citep{Kriek2009} utilizing broad- and medium-band photometry from the 3D-HST survey \citep{ Skelton2014, Momcheva2016} spanning from optical to mid-IR wavelengths. Prior to SED fitting, the photometry in the rest-optical filters were corrected to remove emission line flux from the band \citep[see ][for a more detailed description]{Kriek2015, Reddy2015, Sanders2020}. This procedure fits stellar population models from \citet{Conroy2009} to the data and assumed a \citet{Calzetti2000} dust reddening curve and a \citet{Chabrier2003} IMF.  Additionally, these models assumed a `delayed-$\tau$' star-formation history of the form $\textrm{SFR}(t)\sim t \times e^{-t/\tau}$.

Individual emission lines were measured by first fitting and subtracting a linear local continuum and then fitting a Gaussian profile to the remaining line flux. The science spectra were corrected for slit losses using the method described in \citet{Kriek2015} and \citet{Reddy2015}. Each science spectrum was perturbed by the error spectrum 1000 times and the line was refit with each iteration.  The error was then defined as the $68$th percentile of the width of the resulting distribution.  Multiple lines that lie in close proximity were fit simultaneously, such as the case for the $\rm [OII]\lambda\lambda 3726,3729$ doublet and the $\rm [NII]\lambda\lambda6548,6584+H\alpha$ lines which were fit with a double and triple Gaussian respectively.  After the lines were fit, the galaxy redshift was assigned based on the emission line with the highest SNR, typically H$\alpha$ or [OIII]. Based on the stellar-absorption corrected Balmer emission lines from the best-fit stellar population synthesis model, we used the ratio of Balmer lines to estimate the nebular extinction, and, accordingly the dust-corrected Balmer emission-line fluxes \citep{Reddy2015, Shivaei2020}. For these calculations we assumed the Milky Way dust extinction curve \citep{Cardelli1989}.  The SFR for each galaxy in our sample was calculated based on H$\alpha$, and using the calibration defined by \citet{Hao2011} that has been adjusted to use the \citet{Chabrier2003} IMF.  Figure~\ref{fig:sample} (right) shows the distribution of stellar mass and SFR of our sample.

\subsection{The $z\sim1.5$ Sample}
\label{sec:samples}
We selected galaxies from the MOSDEF survey that lie in the lowest targeted redshift range.  In total, this selection consisted of 314 galaxies with redshifts in $1.37 \le z \le 1.70$.  We then removed galaxies from the sample with stellar masses below a cutoff of $\log(M_*/\rm M_{\odot})=9.0$ as the sample is incomplete for such objects.  Next, objects identified as AGN based on X-ray or rest-frame near-infrared properties were removed from the sample \citep{Coil2015, Azadi2017, Azadi2018, Leung2019}. We further removed galaxies with $\log(\rm [NII]/H\alpha)>-0.3$ as star formation is likely not their primary source of ionizing flux. These criteria resulted in a sample of 285 galaxies with a median stellar mass of $\log(M_*/\rm M_{\odot})=9.97$. We defined a number of subsamples composed of galaxies with different emission-line detection requirements for use in different analyses. Table~\ref{tab:samplenumbers} provides an overview of these different subsamples.  In order to facilitate the creation of composite spectra, the first sample we established necessitates the measurement of at least H$\alpha$ at $>3\sigma$. This requirement resulted in a sample of 238 galaxies from which we constructed composite spectra with wavelength coverage of H$\alpha$ and [NII]. The median stellar mass of these galaxies with H$\alpha$ detected at $>3\sigma$ was comparable to that of the full MOSDEF $z\sim1.5$ sample. We established a sample of galaxies with a $>3\sigma$ detection in H$\alpha$ in addition to coverage of [NII], H$\beta$, and [OIII] totalling 218 galaxies with a median stellar mass of $\log(M_*/\rm M_{\odot})=9.88$. The coverage of these strong lines allowed for the estimate of the median dust-corrected SFR of composite spectra.  In addition, we constructed a sample with $>3\sigma$ detection in both H$\alpha$ and H$\beta$, so that a robust dust-corrected SFR could be estimated for individual galaxies. This sample, comprising 129 galaxies, has a median stellar mass of $\log(M_*/\rm M_{\odot})=9.89$ and median SFR of $\log(\rm SFR/\rm M_{\odot}yr^{-1})=1.10$. Finally, in order to allow for the analysis of the N2 and O3N2 strong-line ratios of individual galaxies, we defined one subsample requiring the detection of $\rm [NII]\textrm{ and }H\alpha > 3\sigma$, and one subsample requiring the detection of H$\alpha$, [NII], H$\beta$, and [OIII]$ > 3\sigma$, comprising 138 and 73 galaxies respectively.  In total, requiring detections of these different lines does not significantly alter the median galaxy properties compared to when all galaxies are included.

\subsection{Composite Spectra}

In order to reduce the impact of any biases incurred by only investigating galaxies with detections in all the available lines, we constructed several composite spectra that include objects for which individual lines could not be measured with high fidelity. A full description of the stacking methodology is presented in \citet{Sanders2018}, however a brief discussion is provided here for convenience. Stacking a subset of galaxies in our sample first required that all objects have coverage of the desired emission lines. To compute the composite spectrum, each individual spectrum contained in the stack was first shifted into the rest frame and converted to luminosity density space based on its measured systemic redshift.   The individual spectra were then normalized based on their H$\alpha$ luminosities and  interpolated onto a common wavelength grid sampled at the median redshift of the galaxies.  Emission line fluxes of composite spectra were measured using the previously described method, however the uncertainties in these measurements were computed using a bootstrap Monte-Carlo method.  First, the quantity that we constructed our bins from (e.g., stellar mass) was perturbed and the galaxies in each bins were redefined.  Then, each individual spectrum was perturbed based on its error spectrum, and the resulting spectra were stacked and the stacked emission lines were remeasured. After we repeated this process 100 times, the uncertainty in the emission-line fluxes was then defined by the inner $68$th percentile of the resulting distribution.

\section{Results}
\label{sec:results}

\begin{table*}
\begin{center}
\renewcommand{\arraystretch}{1.4}
\begin{tabular}{lrrc}
\toprule
    & $\log(\rm [NII]/H\alpha)_0$ & $\log(M_0/\rm M_{\odot})$ & $\gamma$  \\
\midrule
 $z\sim1.5$ This work & $-0.42\pm0.05$ & $9.97\pm0.11$ & $-1.08\pm0.18$  \\
 $z\sim2.3$ \citet{Sanders2018} & $-0.41\pm0.06$ & $10.29\pm0.15$ & $-0.91\pm0.18$  \\

 \bottomrule
 \end{tabular}
 \end{center}
 \caption{Best-fit coefficients of the fit to [NII]/H$\alpha$ as a function of stellar mass, based on stacked spectra.}
\label{tab:N2fits}
\end{table*}

\subsection{Strong-line ratios at $z\sim1.5$}
In this section we present measurements of the strong-line ratios of galaxies in our $z\sim1.5$ MOSDEF sample, and compare them to the observed line ratios in local galaxies from Sloan Digital Sky Survey \citep[SDSS;][]{Abazajian2009}, as well as stacks of galaxies at $z\sim2.3$ from the MOSDEF survey \citep{Sanders2018}. The local comparison sample comes from measurements of $z\sim0$ star-forming galaxy composite spectra from \citet{AM13}. The composites used here are constructed in bins of stellar mass separated by 0.1 dex, and we limit the masses to $9.0 \le \log(M_*/\rm M_{\odot}) \le 11.0$ in order to better match our $z\sim1.5$ data. Based on the available observed wavelength ranges covered for the $z\sim1.5$ galaxies as part of the MOSDEF survey, we analyze the N2 and O3N2 line ratios. Using empirical calibrations one can utilize these line ratios in order to understand the gas-phase metallicities within galaxies. This discussion is presented in Section~\ref{sec:disc} below.

\begin{figure*}
    \centering
    \includegraphics[width=1.0\linewidth]{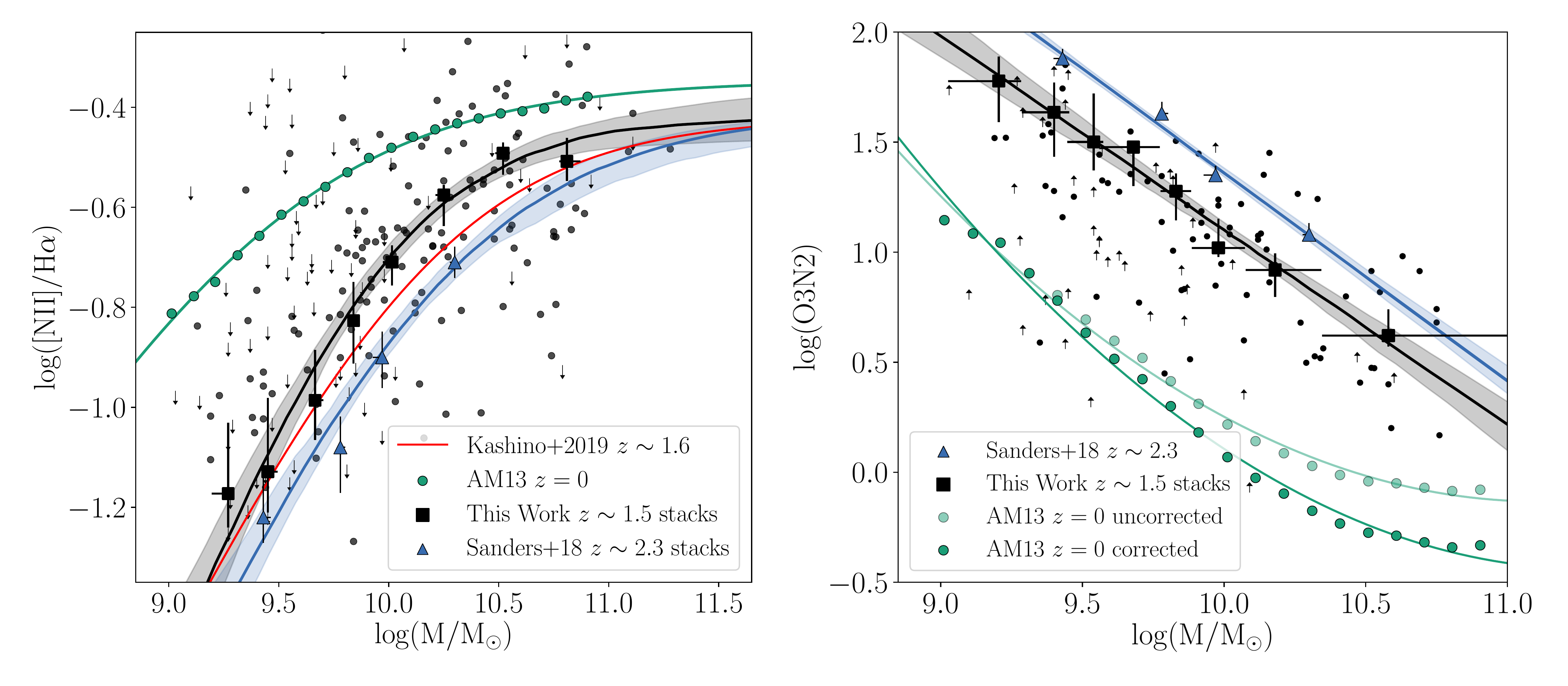}
    \caption{\textit{Left: }$\log(\rm [NII]/H\alpha)$ ratio as a function of stellar mass.  Small black circles show the positions of individual measurements from galaxies in the $z\sim1.5$ MOSDEF sample.  Arrows define the $3\sigma$ upper limit for galaxies that do not have a significant detection in [NII]. Large symbols show measurements from stacked spectra in bins of stellar mass for $z=0$ \citep{AM13}, $z\sim1.5$ (This work), and $z\sim2.3$ \citep{Sanders2018} galaxies in green, black, and blue respectively. The stacked points at each redshift are fit using equation~\ref{eqn:equation}, and the $1\sigma$ uncertainty envelope is displayed for the two high-redshift sample. The best-fit relation from \citet{Kashino2019} is shown in red for comparison. \textit{Right: }Same as the left panel but for the O3N2 emission line ratio. The $z=0$ stacks that are shown have been corrected for the influence of DIG using the method prescribed by \citet{Sanders2017}.}
    \label{fig:allratios}
\end{figure*}

The [NII]/H$\alpha$ ratio is one of the most commonly used probes for investigating properties of the ISM within galaxies. The [NII]/H$\alpha$ ratio is strongly dependent on the ionization parameter and, due to the anticorrelation of metallicity and ionization parameter \citep{PerezMontero2014}, is a useful indicator of the gas-phase oxygen abundance.  However, this ratio is also sensitive to the nitrogen abundance, N/H, and therefore indirectly to O/H due to the N/O to O/H relation.  While the locally established metallicity calibrations that use this line ratio incorporate the N/O to O/H relation, which has been extensively studied in the local universe \citep[e.g.,][]{Pilyugin2012}, using the same calibration in high-redshift galaxies implies this N/O scaling remains consistent at high redshift, where it is less well characterized.  One advantage of this line ratio is the proximity of [NII] and H$\alpha$ in wavelength, meaning that there is very little differential dust attenuation between its constituent lines, and a robust dust correction is not required.  In addition, these two lines can be observed for galaxies that fall into several distinct redshift windows, where the lines lie within regions atmospheric transmission, allowing a direct study of its evolution through cosmic time.   

Figure~\ref{fig:allratios}(left) displays the [NII]/H$\alpha$ ratio of individual galaxies in our $z\sim1.5$ sample, as well as measurements of composite spectra constructed for bins of stellar mass for the 238 galaxies that have a $>3\sigma$ detection in H$\alpha$. Properties of the $M_*$ bins for our $z\sim1.5$ sample are given in Tables~\ref{tab:Mstacks}.  This figure also shows these line ratios for stacks of $z\sim0$ SDSS galaxies based on stellar mass \citep{AM13}, and stacks of $z\sim2.3$ galaxies from the MOSDEF survey \citep{Sanders2018}. The best-fit relation of $z\sim1.6$ stacks from the FMOS-COSMOS survey is displayed in red for comparison \citep{Kashino2019}. As with the relation in the local universe and at $z\sim2.3$, the [NII]/H$\alpha$ ratios of our $z\sim1.5$ stacks increase with increasing stellar mass.  In more detail, at low masses the $z\sim1.5$ stacks follow a power law with slope $\gamma=-1.08$, similar to that at $z=0$ and $z\sim2.3$. At high mass ($\gtrsim 10^{10.3} M_{\odot}$), the relation begins to saturate to a value of [NII]/H$\alpha=-0.5$.  This behavior is also apparent in the stacks of local galaxies as [NII]/H$\alpha$ saturates for star-forming galaxies, however this flattening starts occurring at lower stellar mass at $z\sim0$. A flattening of $\log(\rm [NII]/H\alpha)$ at high masses is not observed in the highest redshift bin (i.e., $z\sim2.3$) considered here, which could be due in part to either the sample not having good coverage at these masses, or an increase of the characteristic turnover mass toward high redshift \citep{Zahid2014}. We fit the stacked points at all three epochs using the functional form of:

\begin{equation}
    \log([\textrm{NII}]/\textrm{H}\alpha) =  \log([\textrm{NII}]/\textrm{H}\alpha)_0 - \log[1+(M_*/M_0)^{\gamma}]
    \label{eqn:equation}
\end{equation}

where $\log(\rm [NII]/H\alpha)_0$ is the asymptotic value at high mass, $M_0$ is the turnover mass, and $\gamma$ is the low-mass power-law slope. This equation is of the same form presented by \citet{Moustakas2011} originally for the O/H vs. $M_*$ relation, and used by \citet{Kashino2019} to parameterize the N2 vs. $M_*$ relation at $z\sim1.6$. Because $\log(\rm [NII]/H\alpha$) and $\log(\rm O/H$) are often related linearly, this equation is appropriate to describe both O/H and N2 as a function of stellar mass.  In addition, this equation is preferred over a power law for the stacks of local galaxies, as well as composite spectra of our $z\sim1.5$ sample which both saturate towards higher masses. At $z\sim1.5$, if the highest mass bins are excluded and a power law is fit to the data, we recover the low mass slope of the above equation, $\gamma$. As mentioned above, the $z\sim2.3$ stacks do not show this characteristic flattening at high masses, however, we show the $z\sim2.3$ stacks fit with the same functional form for consistency. The best-fit parameters for the $z\sim1.5$ and $z\sim2.3$ stacks are given in Table~\ref{tab:N2fits}. The $z\sim1.6$ stacks of \citet{Kashino2019} have lower [NII]/H$\alpha$ at fixed stellar mass compared to our measurements at $z\sim1.5$, however, the best-fit parameters agree within $1\sigma$ for both samples. 

\begin{table}
\begin{center}
\renewcommand{\arraystretch}{1.4}
\begin{tabular}{lcr}
\toprule
    & Slope  & Intercept  \\
\midrule
 $z\sim1.5$ This Work & $-0.88\pm0.10$ & $9.91\pm0.99$   \\
 $z\sim2.3$ \citet{Sanders2018} & $-0.94\pm0.05$ & $10.72\pm0.46$   \\

 \bottomrule
 \end{tabular}
 \end{center}
 \caption{Best-fit coefficients for the equation $\log(\textrm{O3N2}) = m \times \log(M_*/\rm M_{\odot}) + b$ based on stacked spectra at high redshift.} 

\label{tab:O3N2fits}
\end{table}

We also investigate how an additional line ratio, O3N2, varies with galaxy properties.  As with $\log(\rm [NII]/H\alpha)$, this line ratio is constructed so that there is minimal effect of a differential dust correction. The O3N2 ratio is sensitive to both the excitation and metallicity of the ISM, making it effective for measuring oxygen abundances. However, in contrast to the N2 line ratio, O3N2 is anticorrelated with oxygen abundance, and does not saturate for star-forming galaxies. In addition, the same range of gas-phase metallicities corresponds to a larger range of O3N2 compared to N2 meaning uncertainties in the line ratio translate into smaller metallicity uncertainties when using O3N2.

Figure~\ref{fig:allratios}(right) displays the O3N2 ratios as a function of stellar mass for the galaxies in our $z\sim1.5$ sample (black points). The $z\sim1.5$ O3N2 composite measurements, which comprise galaxies with a $>3\sigma$ detection of H$\alpha$ and coverage of [OIII] and H$\beta$, show a clear anti-correlation with stellar mass, which is also observed in the $z\sim2.3$ and local universe stacks.  In this figure we also display limits for galaxies that do not have detections in all components of O3N2, but do have a $>3\sigma$ detection in H$\alpha$.  These limits become more prevalent in galaxies at lower masses. We create stacks that include galaxies with limits using the methods described above to lessen biases induced when only galaxies with $>3\sigma$ detections are considered. The stacked values of O3N2 lie directly within the individual measurements at high mass, and slightly above most of the individual detections at low masses, which is compatible with the trend of lower limits which appear in our sample in this lower-mass regime.  The stacked points display a clear power-law relation between O3N2 and stellar mass, in contrast to N2 which flattens out at high masses.  However, because the stellar mass bins are constructed to contain an equal number of galaxies, the highest stellar mass bin covers a large range of stellar masses ($\sim0.5$ dex) which may smooth out any apparent variations in the relation.  The best-fit linear function coefficients are given in Table~\ref{tab:O3N2fits}.  The slope of this fit is consistent with that of the $z\sim2.3$ stacks, however it is offset towards lower O3N2 by $\sim0.25 \rm\  dex$.

\subsection{Secondary dependence of the strong-line ratios}
\label{sec:fmr}

\begin{figure}
    \centering
    \includegraphics[width=1.0\linewidth]{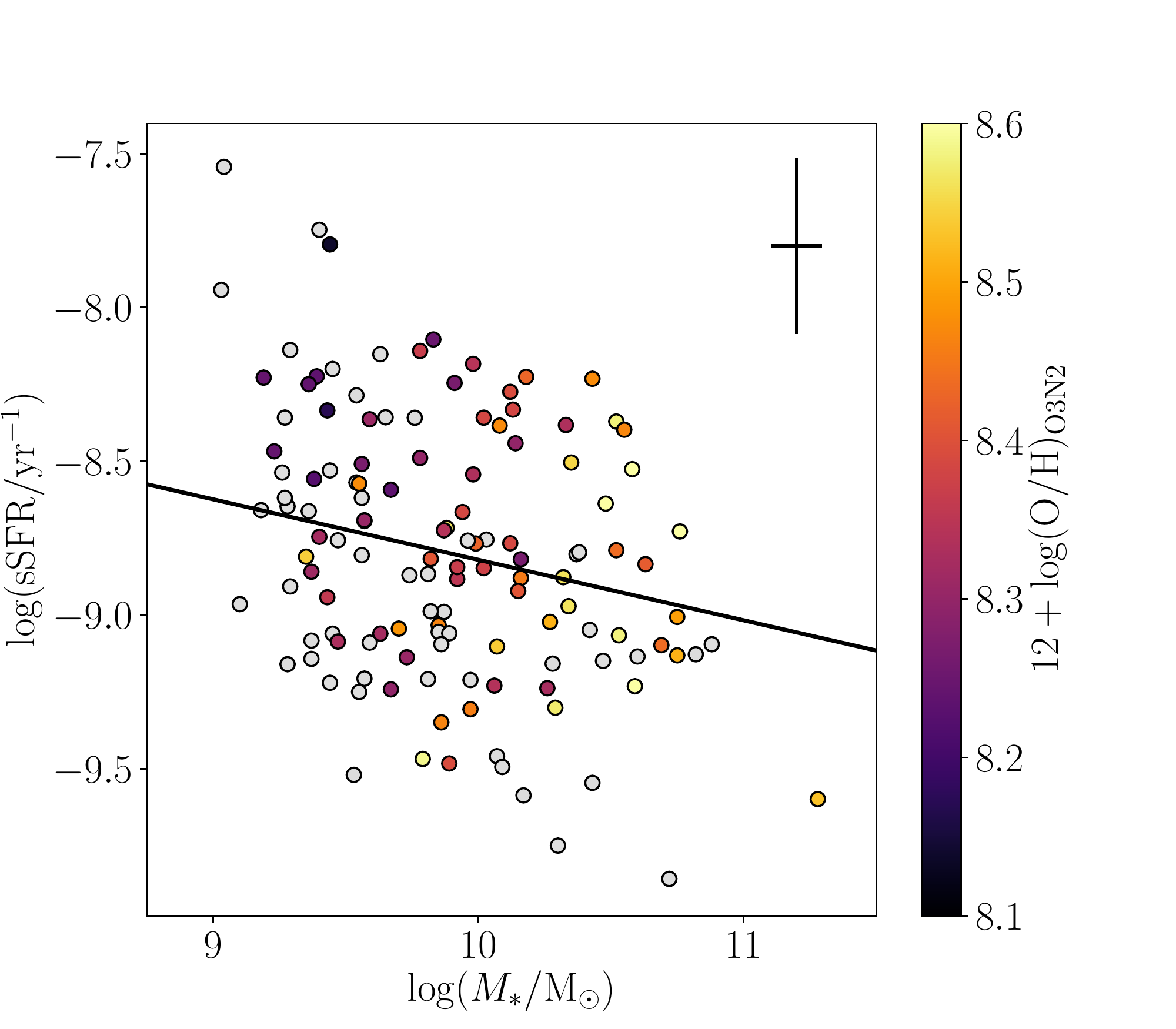}
    \caption{The sSFR vs. $M_*$ relation measured for galaxies in our $z\sim 1.5$ MOSDEF sample.  Points that have a $>3\sigma$ detection in H$\alpha$, [NII], H$\beta$, and [OIII] are color-coded based on their oxygen abundance calculated using the \citet{PP04} O3N2 calibration. Grey circles show properties of individual galaxies without such detections.  The black line shows the best-fit relation to the stacked points.}
    \label{fig:ssfr}
\end{figure}

\begin{figure}
    \centering
    \includegraphics[width=1.0\linewidth]{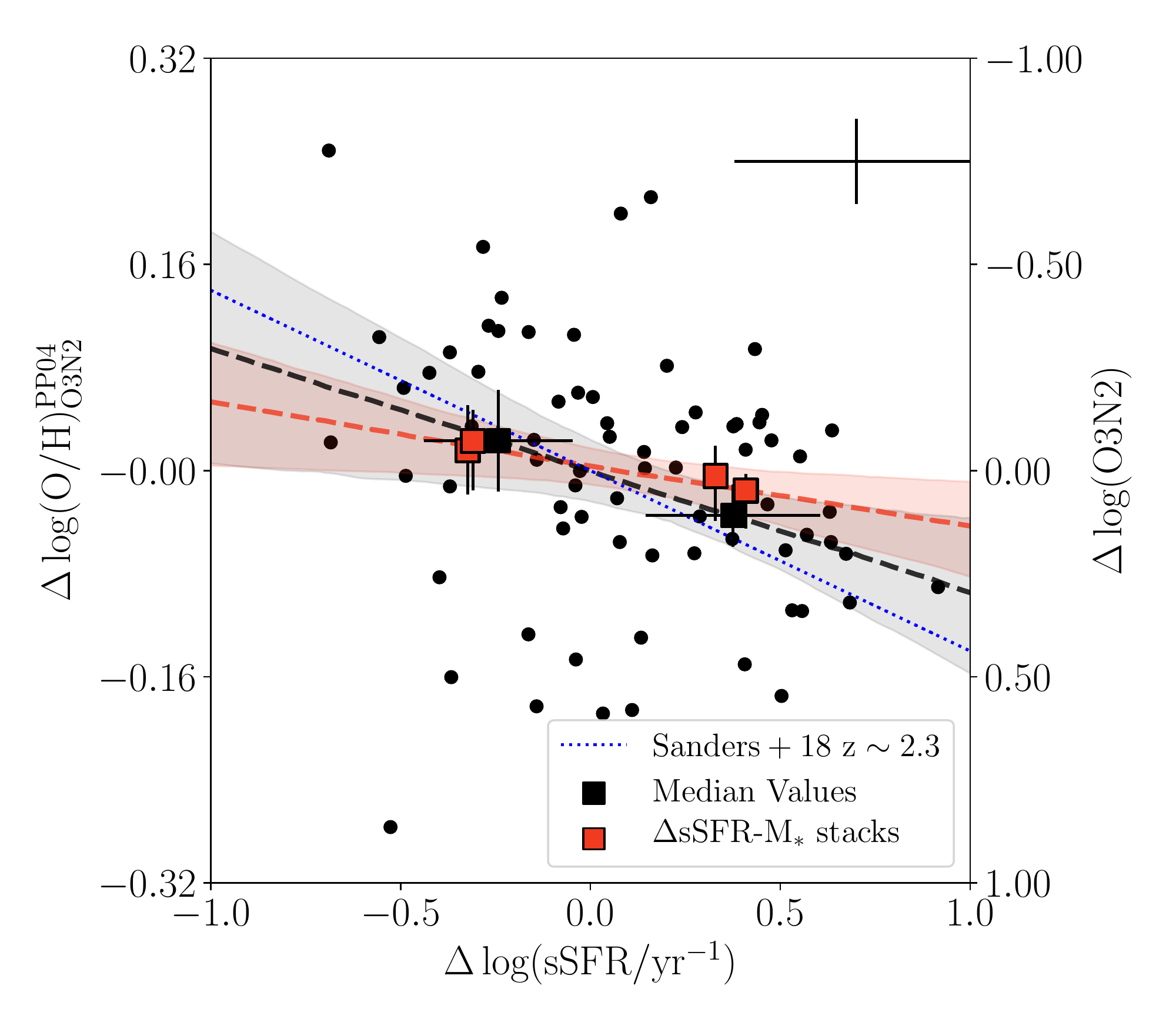}
    \caption{Excess $\log(\rm O/H)$ vs. excess $\log(\rm sSFR)$. The oxygen abundances are calculated using the PP04 calibration of the O3N2 line ratio. Small black points are measurements for individual galaxies in our $z\sim1.5$ sample that have $>3\sigma$ detections in [NII], H$\alpha$, [OIII], and H$\beta$. The large black points are median values calculated from the sample of individual points that has been divided into two bins of $\Delta \log(\rm sSFR)$. The red points are measurements from stacked spectra composed of galaxies in the full sample in bins of stellar mass and sSFR as described in Section~\ref{sec:fmr}.  The dashed black and red lines are the best-fit lines to the median and $M_*$-sSFR stacked points respectively.  The colored shaded regions show the $1\sigma$ envelope for the parameters of the best fit relations. Both best-fit relations show a positive correlation, and are consistent with each other to within $1\sigma$, suggesting that this correlation is real. The blue dotted line shows the best fit relation of $z\sim2.3$ MOSDEF galaxies from \citet{Sanders2018}.}
    \label{fig:O3N2FMR}
\end{figure}

The previous section demonstrates how strong-line ratios that are sensitive to the gas-phase metallicity are closely related to galaxy stellar mass.  A more detailed analysis of galaxies in the local universe has revealed a secondary dependence of the gas-phase metallicity on the SFR resulting in a $M_*$-SFR-Z relationship, known as the FMR \citep{Mannucci2010}. Evidence also suggests  that the FMR is present at high redshift \citep{Cullen2014, Zahid2014b,Kashino2017, Sanders2018, Sanders2020}. In this section we investigate evidence for the FMR in our $z\sim1.5$ data using two methods.

One piece of evidence for the presence of the FMR in our data manifests is significant correlation between the scatter of galaxies about the MZR, and the mean $M_*$-specific SFR (sSFR$\equiv (\rm SFR /M_{\odot} yr^{-1})/(M_*/M_{\odot})$) relation (Figure~\ref{fig:ssfr}). In particular, galaxies that are offset toward higher sSFR at fixed $M_*$ will have lower metallicities, and galaxies offset toward lower sSFR will higher metallicities.  We search for this correlated scatter among both the detections of individual galaxies, as well as stacks comprising our full $z\sim1.5$ sample.  While an ideal determination of the FMR primarily uses the detections of individual galaxies, this requires detections of all lines used to estimate the gas-phase metallicity (in this case O3N2), which may introduce biases.  We construct stacks of galaxies by dividing the sample of objects with $>3\sigma$ detections in H$\alpha$ and H$\beta$ into those that fall above and below the mean $M_*$-sSFR relation. We further divide each subsample into two equal sized bins based on $M_*$, resulting in a total of four composites.  We then measure O3N2 from the resulting composite spectrum, and define the stack $M_*$ and sSFR as the medians of galaxies within each stack.

Figure~\ref{fig:O3N2FMR} displays the offset from the mean $M_*$-$\log(\rm O/H)$ relation ($\Delta \log(\rm O/H)_{\rm O3N2}$), against the offset from the mean $M_*$-sSFR relation ($\Delta \rm sSFR$) as shown in Figure~\ref{fig:ssfr}. We use the O3N2 line ratio to calculate the oxygen abundance as it shows a stronger correlation with $\Delta \rm sSFR$ compared to using N2 in $z\sim2.3$ galaxies \citep{Sanders2018} and, given our smaller sample, therefore is a more sensitive probe of the FMR at $z\sim1.5$. First, we can consider the correlation between these two quantities using only measurements from individual galaxies.  We split the individual detections into two equal bins of $\Delta \rm sSFR$ and plot the median values of these bins in black.  We recover a slope of $-0.09\pm0.06$ which is consistent with an anticorrelation at  the $1.5\sigma$ level.  We estimated the uncertainty of the slope by perturbing the individual galaxy spectra by their uncertainties, recalculating the medians, and then finding the best-fit line. This slope is shallower than that of best-fit relation of $z\sim2.3$ galaxies of $-0.14 \pm 0.034$ from \citet{Sanders2018}, but consistent within $1\sigma$. Analysis of the individual detections using a Spearman correlation test reveals a correlation coefficient of $r_s=0.317$ with a p-value of $0.006$.  This suggests a  rather weak, although significant correlation between these two parameters. However, since this estimation only includes those galaxies with $>3\sigma$ detections in O3N2, many galaxies in our sample are not included, and a more robust measurement of this trend includes composite spectra of all galaxies in our sample.   We test for this correlation using the previously described $M_*$-sSFR stacks, which comprise all galaxies in our sample with a $3\sigma$ detection in H$\alpha$ and H$\beta$.  These stacks are displayed in red in Figure~\ref{fig:O3N2FMR}.  We find a slope of $-0.05\pm0.04$ when fitting a linear function to the stacked points.  Again, this slope is consistent with an anticorrelation at the $1\sigma$ level, and consistent with the slope found for the median points of the individual galaxies. While the value of the slope between these quantities likely varies due to sample differences, these results suggest that there is a real correlation between the scatter of these two relations, providing evidence for the existence of the $\rm M_*$-SFR-Z relation at $z\sim 1.5$. However, a larger sample is needed to establish this anticorrelation with greater significance.

\begin{figure*}
    \centering
    \includegraphics[width=1.0\linewidth]{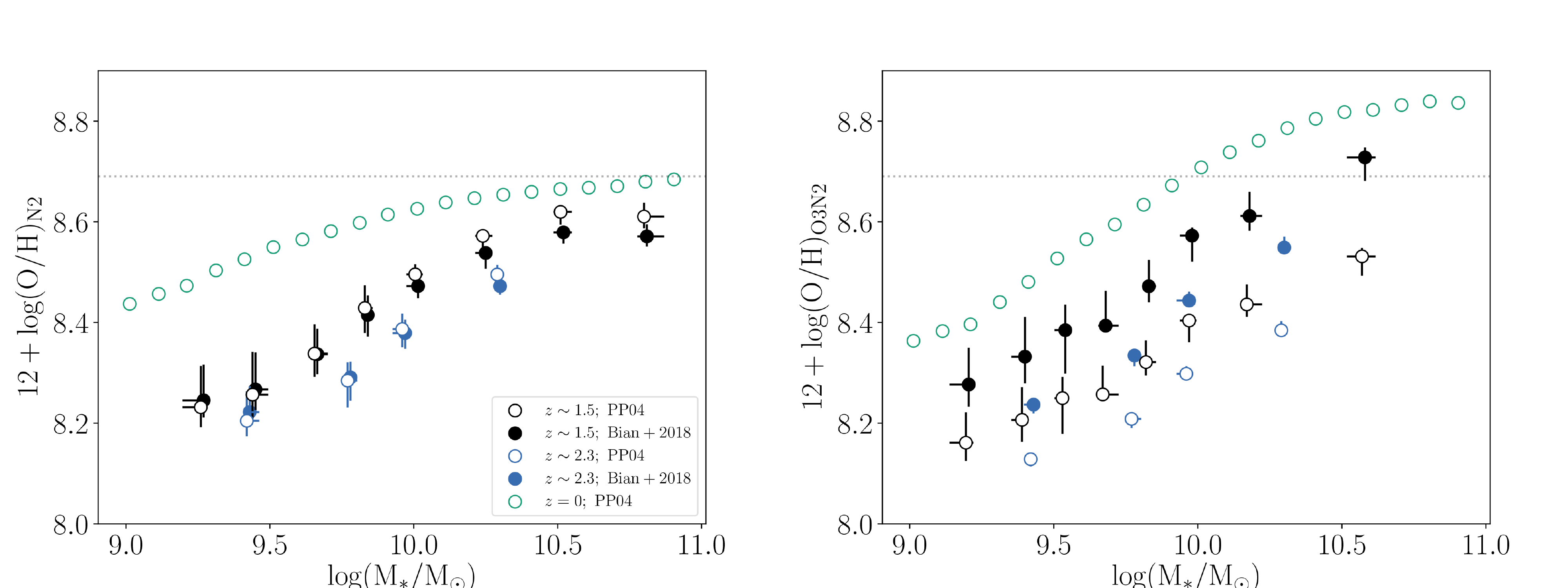}
    \caption{\textit{Left}: MZR at $z=0$, $z\sim1.5$, and $z\sim2.3$ shown as green, black, and blue points respectively. The metallicities in this panel are all calculated using calibrations that utilize the [NII]/H$\alpha$ line ratio. Filled markers use the metallicity calibrations from \citetalias{PP04}, and open markers are calculated based on the calibrations from \citetalias{Bian2018}.  The \citetalias{Bian2018} calibration does not cover the range of metallicity spanned by the $z=0$ galaxies, therefore we do not plot the local metallicities calculated using this calibration. For reference, solar metallicity is shown as the grey dotted line. \textit{Right}: Same as (left) but for metallicities calculated using calibrations based on the O3N2 line ratio.}
    \label{fig:mzrs}
\end{figure*}

\begin{figure*}
    \centering
    \includegraphics[width=1.0\linewidth]{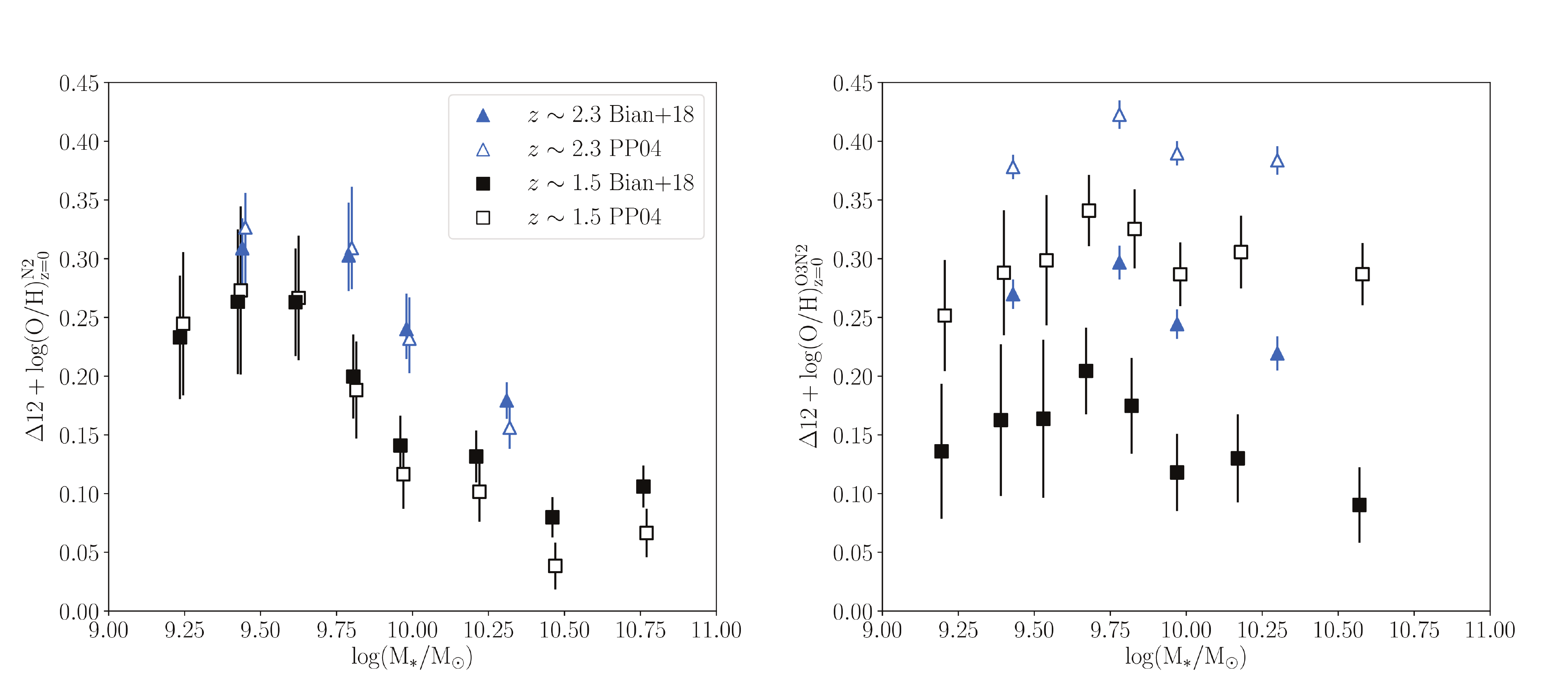}
    \caption{\textit{Left}: Offset of $\log(\rm O/H)_{\rm [NII]/H\alpha})$ of high redshift galaxy stacks from the local galaxy sequence as a function of stellar mass. The black squares and the blue triangles are for the $z\sim1.5$ and $z\sim2.3$ galaxies respectively, and the filled and empty symbols show metallicities calculated using \citetalias{Bian2018} and PP04 calibrations respectively. At $z\sim1.5$ there is an $\log(\rm O/H)_{\rm N2}$ ($\log(\rm [NII]/H\alpha)$) offset of $\sim0.25\rm\  dex$ (0.45 dex) at the lowest mass stack, down to an offset of $\sim0.05\rm\ dex$ (0.10 dex) at the highest masses calculated using the PP04 calibration. The $z\sim2.3$ stacks follow the same trend, but are $\sim0.1$ dex higher in N2 and $\sim0.05$ dex higher in $\log(\rm O/H)$ on average. \textit{Right}: Same as (left) but using the O3N2 emission line ratio. The offset from the local sequence when using this line ratio does not strongly depend on stellar mass.}
    \label{fig:deltas}
\end{figure*}

\section{Discussion}
\label{sec:disc}
\subsection{Evolution of the MZR}
In the previous section we described how different observed strong-line ratios depend on their host galaxy properties.  We now discuss how these strong-line ratios are used to calculate the oxygen abundance in these galaxies.

We use calibrations to estimate the oxygen abundance based on the strong-line ratios N2 and O3N2. In particular, we consider the calibrations from \citet{PP04} and \citet{Bian2018}, hereafter referred to as PP04 and B18, respectively, as they represent different methodologies to produce oxygen abundance measures.  The PP04 calibrations for both N2 and O3N2 are based on direct-method metallicities of HII regions in the local universe. While this calibration has been shown to be appropriate for measuring the metallicity of galaxies in the local universe, changing physical conditions in high-redshift galaxies suggest that this calibration may not be able to accurately reproduce gas-phase metallicities at high redshift \citep{Steidel2016,Bian2018,Sanders2018, Sanders2019,  Topping2020, Topping2020b,  Runco2020}.  In contrast, the calibrations produced by B18 are constructed based on local galaxies that share certain emission line properties as those at high redshift, i.e., that lie in the same place on the  [OIII]/H$\beta$ vs. [NII]/H$\alpha$ BPT diagram.  This method ensures that the empirical metallicity calibrations that are applicable to high-redshift galaxies were constructed using measurements of galaxies that share similar emission-line ratio properties to those at high redshift.  However, the location on the BPT diagram may not be sufficient in order to classify galaxies based on their physical properties \citep{Topping2020, Topping2020b, Runco2020}.  In addition, the region of the BPT diagram used to classify the local analogs of high-redshift galaxies to construct the B18 calibration was defined by the Keck Baryonic Structure Survey (KBSS) $z\sim2.3$ sample \citep{Steidel2014}; the MOSDEF survey, based on different selection criteria, yields a $z\sim2.3$ locus displaying a smaller offset from $z=0$ galaxies on the BPT diagram \citep{Shapley2019, Sanders2020}.  

Figure~\ref{fig:mzrs} shows the MZR calculated using both the N2 and O3N2 calibrations for our stacks of $z\sim1.5$ galaxies, the stacks of galaxies at $z\sim2.3$ \citep{Sanders2018}, and local galaxies \citep{AM13}.  This figure displays the metallicities of the two high-redshift stacked spectra calculated using both the \citetalias{PP04} and \citetalias{Bian2018} calibrations.  The N2-based metallicities for these calibrations only differ at high masses ($>10^{10}\rm  M_{\odot}$) by $\sim0.05$ dex, and are consistent at low masses.  However, metallicities calculated based on the O3N2 line ratio differ by constant $\Delta\rm log(\rm O/H)\sim0.1$ dex across the mass range we investigate.  Figure~\ref{fig:deltas} displays the difference between the high-redshift metallicities and the metallicities of local galaxies at fixed $M_*$.  This figure illustrates two trends: the inferred redshift evolution in the MZR between $z\sim1.5$ and $z=0$ differs depending on which line ratios are used to calculate the metallicities, as well as the metallicity offset.  For the N2 ratio, at the lowest masses the difference between $z=0$ galaxies and $z\sim1.5$ galaxies is $\Delta\rm log(\rm [NII]/H\alpha)\sim0.45$ dex, which decreases to $\Delta\rm log(\rm [NII]/H\alpha)\sim0.10$ dex at the highest masses.  The $z\sim2.3$ stacks show a similar trend, but is offset towards lower N2 compared to the $z\sim 1.5$ stacks by $\Delta\rm log(\rm [NII]/H\alpha)\sim0.10$ dex on average.  In metallicity space based on the N2 calibration, at fixed stellar mass the $z\sim1.5$ composites are offset to lower metallicity by $\Delta\rm log(\rm O/H)\sim0.25$ dex compared to local galaxies for the lowest masses which narrows to an offset of $\Delta\rm log(\rm O/H)\sim0.05$ dex at high mass.  In contrast, the O3N2 ratio exhibits a nearly constant offset of $\Delta\rm log(\rm O/H)\sim0.30$ dex compared to z=0 galaxies across the mass range explored here when using the \citetalias{PP04} calibration. Similarly, the $z\sim2.3$ stacks show a similar constant offset as a function of mass, with O3N2 ratios that are $\Delta\rm O3N2\sim0.2$ dex higher on average when compared to $z\sim1.5$ galaxies. Accordingly, we find that the low-mass slope of the MZR does not evolve significantly when using O3N2-based metallicity calibrations, and the slope does exhibit evolution between $z\sim1.5$ and $z=0$ when using N2. This slope evolution based on N2 is in with conflict with the MZR measured at $z\sim2.3-3.3$ from \citet{Sanders2020}, which used more robust metallicity measurements based on a larger set of rest-optical emission line ratios involving only $\alpha$ elements.  Based on the lack of MZR slope evolution when using O3N2 as a metallicity indicator (where the lack of MZR slope evolution is in agreement with the results of \citet{Sanders2020}), due in part to the saturation of N2 at high masses \citep{PP04}, we argue that O3N2 provides a more robust metallicity calibration than N2 for our $z\sim1.5$ sample.

The shape and evolution of the MZR has been explained using both cosmological simulations and analytic chemical evolution models that balance the exchange of gas within galaxies and their environments, star formation, and chemical enrichment \citep[see e.g.,][]{Finlator2008, Peeples2011, Dave2012, Lilly2013, Feldmann2015, Ma2016, Dave2017, Torrey2019}.

In the equilibrium or gas-regulator analytic chemical evolution models \citep{Peeples2011, Dave2012, Lilly2013}, the gas-phase metallicity is regulated by the balance of accretion, outflows, and gas dilution, all of which can vary with stellar mass. These studies demonstrated that significant gas accretion and outflows are required to explain the shape and normalization of the MZR at $z=0$.  In the context of these models, the decrease in MZR normalization with increasing redshift can be explained by either larger gas fractions or more metal expulsion via outflows at higher redshifts.

In terms of cosmological simulations of galaxy formation, \citet{Ma2016} investigated the MZR using the FIRE simulation suite, and found MZR evolution that is associated with changes in the gas fraction in galaxies at fixed stellar mass. This result hints to an additional dependence of the MZR on gas mass. Similarly, \citet{Dave2017} utilized the MUFASA simulations to show that at fixed stellar mass, the total gas content of galaxies is higher at earlier epochs. They are also able to reproduce the slope and normalization of the MZR out to $z=2$ at low mass. However, the MZR for $M_*\ge 10^{10}\ \rm M_{\odot}$ galaxies in MUFASA does not agree with observations, possibly due in part to an incorrect implementation of wind recycling, which contributes to setting the MZR at high mass. \citet{Torrey2019} argues using the IllustrisTNG simulations that the normalization of the MZR is not due to changes in the ability of galaxies to retain metals, but is in fact a result of evolving gas fractions within galaxies.  The results from this simulation suggest that galaxies at high redshift are more efficient at retaining their metals. However, the increased gas masses of high-redshift galaxies cause a net dilution of metals, and thus a lower metallicity.

In contrast, \citet{Sanders2020b} applied the models of \citet{Peeples2011} to the observed MZR over a range of redshifts from $z=0-3.3$ using data from the MOSDEF survey.  These authors found that while the slope of the MZR is invariant  out to high redshift, the evolving normalization of the MZR requires a combination of an elevated efficiency of galaxies expelling their metals and increased gas fractions at fixed stellar mass. Our results at $z\sim1.5$ of a constant slope and evolving normalization are consistent with the results of \citet{Sanders2020b}, which suggests a picture in which, as redshift increases, galaxies become more gas rich and winds are more effective at removing metals from the galaxies.

The MZRs measured at $z\sim1.5$ in previous works based on the N2 calibration show a similar trend of decreasing offset from the $z=0$ MZR with increasing mass.  \citet{Yabe2014} measured the MZR from composite spectra comprising 343 star-forming galaxies at $z\sim1.4$.  While these measurements do not probe down to the lowest masses covered by our sample, the MZR at $\log(\rm M_*/M_{\odot})\gtrsim9.5$ is consistent with our finding using N2-based calibrations, with metallicities of $12+\log(\rm O/H)\sim8.45$ at $\log(\rm M_*/M_{\odot})=10.0$ increasing to $12+\log(\rm O/H)\sim8.6$ at $\log(\rm M_*/M_{\odot})=11.0$.  The MZR of $z\sim1.6$ star-forming galaxies from \citet{Zahid2014} similarly probes a mass range of $9.7\le\log(\rm M_*/M_{\odot})\le11.0$ and shows a characteristic turnover at high mass to an oxygen abundance of $12+\log(\rm O/H)\sim8.65$ which is observed in our results. \citet{Kashino2017} uses the calibration of \citet{Maiolino2008} to calculate oxygen abundance from N2 resulting in an MZR that is offset by $\sim0.2-0.3$ dex towards higher O/H at all stellar masses.  However, they find a similar difference between the $z=0$ and $z\sim1.6$ MZRs which ranges from $\sim0.3$ dex at low mass to $\sim0.05$ dex for the highest mass stack.

\begin{figure}
    \centering
    \includegraphics[width=1.0\linewidth]{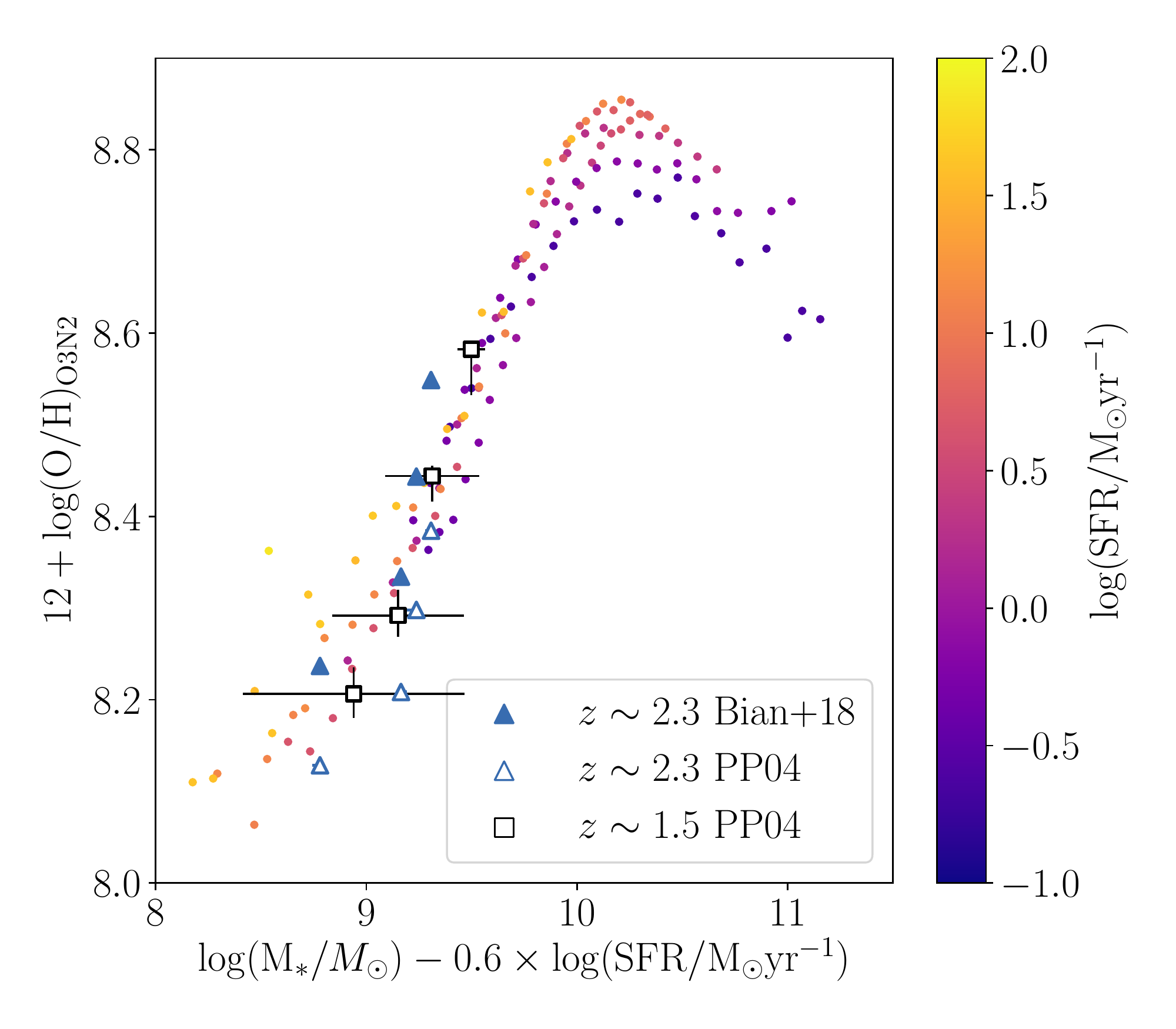}
    \caption{Gas-phase metallicity calculated from the O3N2 line ratio as a function of $\mu=\log(M_*/M_{\odot}) - \alpha \times \log(\rm SFR / M_{\odot} yr^{-1}$ for $\alpha=0.6$.  Colored circles are measurements from composite spectra of local galaxies from \citet{AM13} in bins of 0.1 dex in both stellar mass and SFR. The points are colored based on their SFR.  Black squares are measurements from our $z\sim1.5$ composites constructed in bins of stellar mass with oxygen abundances calculated using the \citetalias{PP04} calibration.  The $z\sim1.5$ measurements lie directly on the sequence traced by the local galaxies.  Blue filled and open triangles show stacks of $z\sim2.3$ metallicities calculated using \citetalias{Bian2018} and \citetalias{PP04} metallicity calibrations respectively.}
    \label{fig:fmr_evo}
\end{figure}
\subsection{Evolution of the FMR}
Several theoretical studies have made predictions for the FMR, relating the stellar mass, metallicity, and SFR of galaxies \citep{Finlator2008, Lilly2013, Ma2016, Dave2017, Torrey2019}.  Many of these studies suggest that the more fundamental relation is one of stellar mass, metallicity, and gas mass, and the FMR arises due to the connection between gas mas and SFR \citep[e.g.,][]{Kennicutt1998}. While the existence of an FMR at high redshift implies that the gas content, and therefore the SFR, is an important driver of chemical evolution, an unchanging FMR throughout cosmic time suggests that galaxies at different times are governed by the same scaling of metal-enriched outflows and gas fractions as a function of stellar mass and SFR, as argued by \citet{Sanders2020b} based on MOSDEF data at $z\sim2.3$ and $z\sim3.3$.

We provide one additional test for the existence of the FMR at $z\sim 1.5$, and compare it to the FMR measured in the local universe.  We project the galaxies in our sample onto the $12+\log(\rm O/H)$ vs. $\mu$ plane, where $\mu=\log(M_*/M_{\odot}) - \alpha \times \log(\rm SFR / M_{\odot} yr^{-1}$) following the method presented in \citet{Mannucci2010}.   At $\alpha=0$, this relation reduces to the MZR, where the lines of constant SFR are distinct from each other.  \citet{Sanders2020} found that a value of $\alpha=0.6$ minimized the scatter among $z\sim0$ composite spectra.  We compare stacks at $z\sim1.5$ and $z\sim2.3$ to the $12+\log(\rm O/H)_{\rm O3N2}$ vs. $\mu$ relation of local galaxies. To make this comparison, we calculated metallicities using calibrations based on the O3N2 line ratio, as it is observed over a larger dynamic range, and is less susceptible to saturation at high masses compared to N2. 

Figure~\ref{fig:fmr_evo} shows the position of $z=0$ galaxies stacked in bins of stellar mass and SFR from \citet{AM13}, which form a fairly tight sequence of increasing O/H as $\mu$ increases, up to a value of $\mu\sim10$, at which point the sequence turns over. However, this regime is outside of the range where the high-redshift galaxies lie.  The metallicities of our $z\sim1.5$ stacks estimated using the \citetalias{PP04} calibration fall along the $12+\log(\rm O/H)$ vs. $\mu$ relation traced by the local galaxies, suggesting that this sequence does not significantly evolve out to $z\sim1.5$.  In terms of the location on the [OIII]/H$\beta$ vs. [NII]/H$\alpha$ BPT diagram, the rest-optical emission line properties of our $z\sim1.5$ stacks display only a slight offset from the sequence of $z\sim0$ star-forming galaxies, whereas the analogs utilized in \citetalias{Bian2018} have significantly higher [NII]/H$\alpha$ at fixed [OIII]/H$\beta$ and are therefore not representative of our $z\sim1.5$ galaxies.  At $z\sim2.3$ the oxygen abundances calculated using the \citetalias{PP04} calibration fall slightly below the local $12+\log(\rm O/H)$ vs. $\mu$ relationship at low mass, however, the highest mass stack is consistent with the local stacks. When using the \citetalias{Bian2018} calibration the measurements of the highest masses fall above the local relation, and the low-mass stacks lie along the local sequence. This shift of $\sim0.1$ dex among the two calibrations illustrates the importance of understanding which calibrations are applicable at high redshift to form an accurate picture of FMR evolution. For $z\sim2.3$ MOSDEF galaxies, the location in the BPT diagram is intermediate between the local SDSS sample, and the $z\sim2.3$ sample from the KBSS survey that \citetalias{Bian2018} attempted to match.  Accordingly, the \citetalias{PP04} and \citetalias{Bian2018} calibrations likely bracket the correct calibration for $z\sim2.3$ MOSDEF galaxies. In addition, both the \citetalias{PP04} and \citetalias{Bian2018} calibrations are based on local galaxies, however directly comparing the emission-line properties of local and high-redshift galaxies may introduce systematics \citep{Sanders2020,  Topping2020, Topping2020b, Runco2020}. There is a clear need for an empirical metallicity calibration for $z\geq1.5$ galaxies based on direct-metallicity measurements of galaxies at the same high redshifts in order to directly compare the normalization of the FMR across different epochs.

\begin{figure*}
    \centering
    \includegraphics[width=1.0\linewidth]{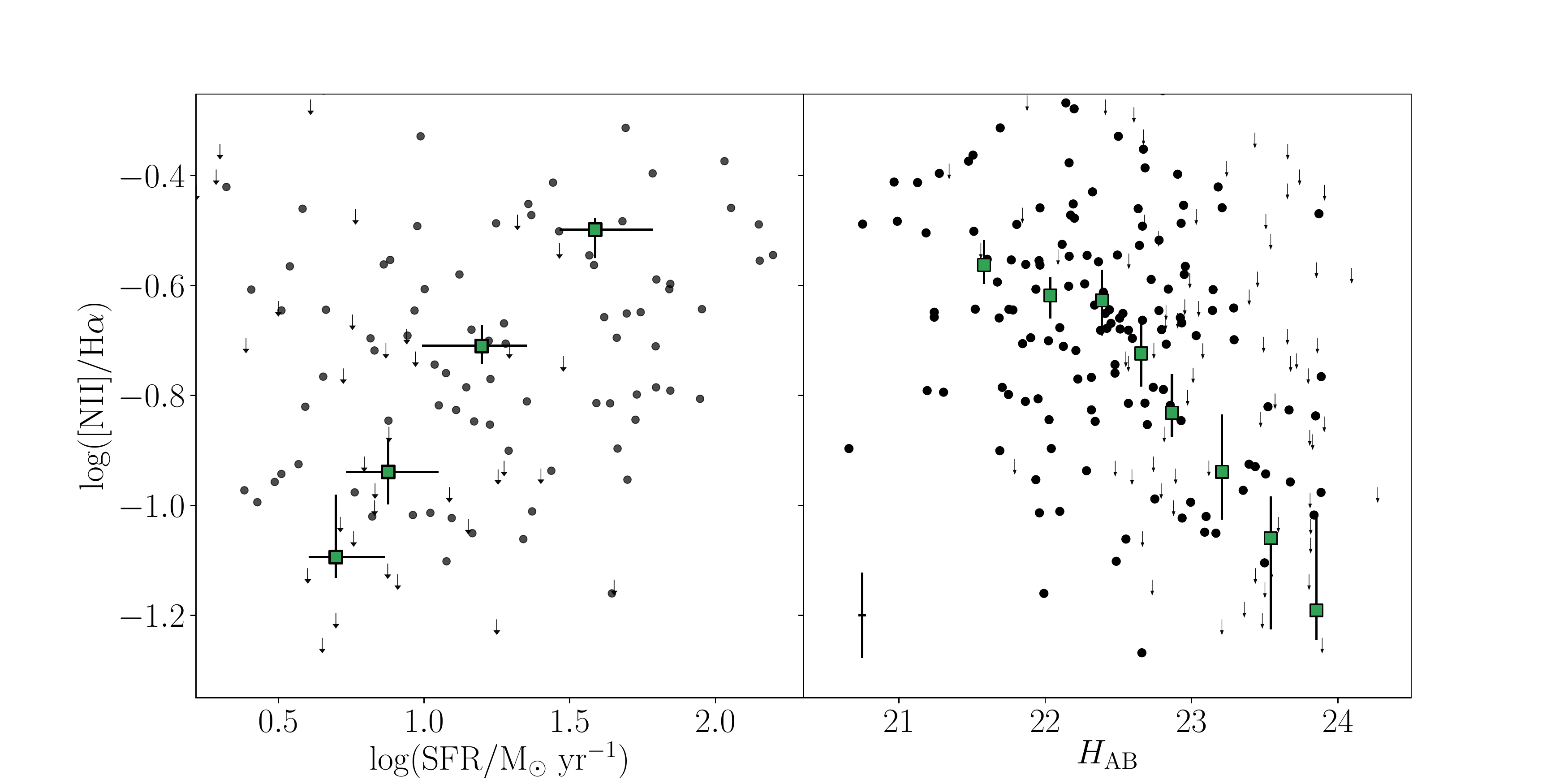}
    \caption{ \textit{Left}:  $\log(\rm [NII]/H\alpha)$ ratio as a function of SFR.  The small points and arrows depict, respectively, measurements and upper limits of individual galaxies in the $z\sim1.5$ sample.  The SFR of galaxies in this figure were calculated based on dust-corrected H$\alpha$, thus requiring a $>3\sigma$ detection in both H$\alpha$ and H$\beta$.  The green square points are [NII]/H$\alpha$ ratios are measurements of stacked spectra in bins of SFR.   \textit{Right}: $\log(\rm [NII]/H\alpha)$ ratio as a function of $H$-band magnitude displayed using the same scheme as in the left panel. These stacks include all galaxies in our $z\sim1.5$ sample, including those without significant H$\beta$ detections.}
    \label{fig:NIIothers}
\end{figure*}

\begin{figure*}
    \centering
    \includegraphics[width=1.0\linewidth]{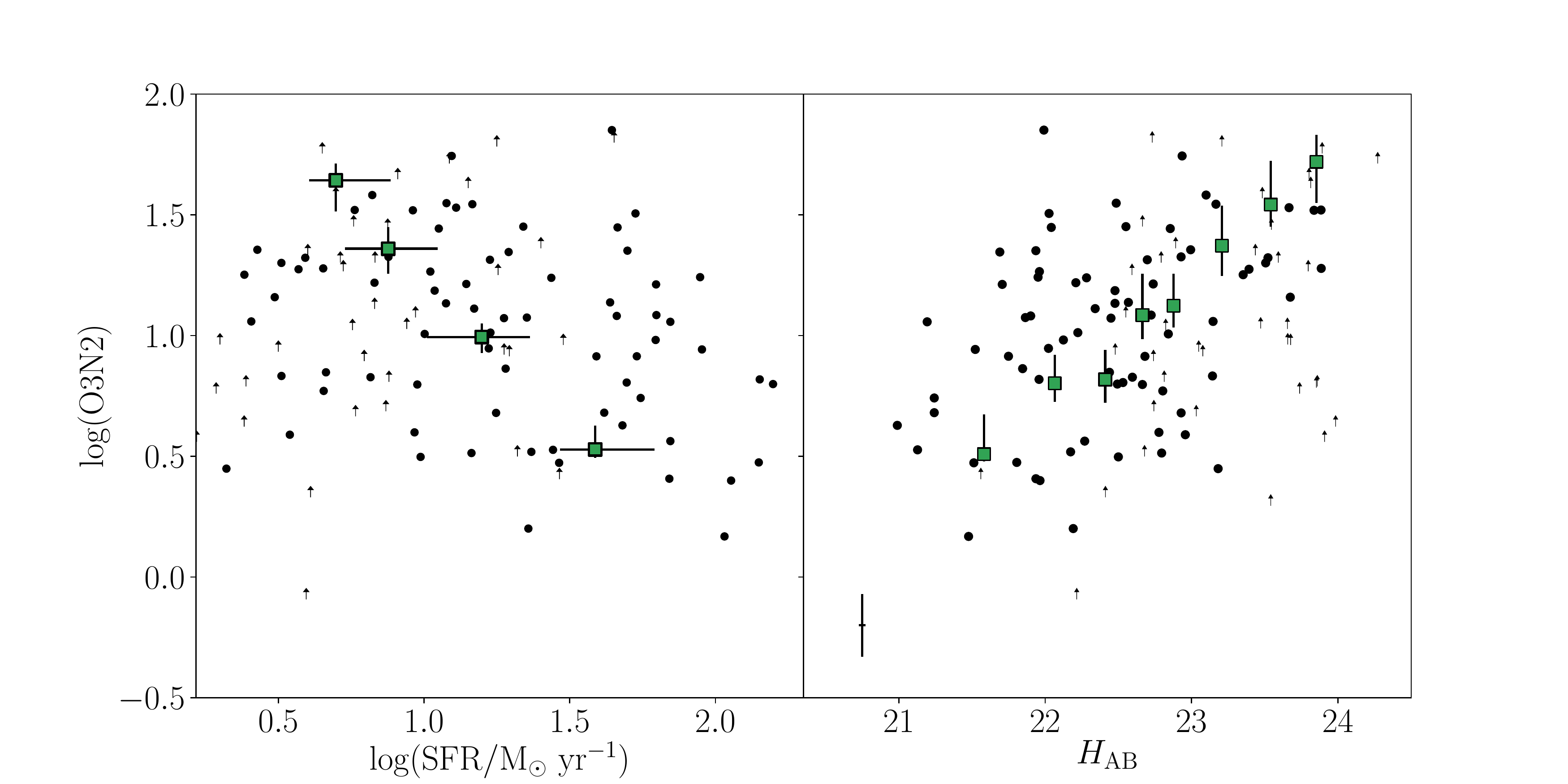}
    \caption{Same as Figure~\ref{fig:NIIothers} but for the O3N2 emission line ratios.}
    \label{fig:o3n2others}
\end{figure*}

\begin{table}
\begin{center}
\renewcommand{\arraystretch}{1.4}
\begin{tabular}{rrrr}
\toprule

    $N_{\rm gal}$ & $\rm SFR_{\rm med}\ [M_{\odot}\rm yr^{-1}]$ $^{\rm a}$ & $\log(\rm [NII]/H\alpha)$ $^{\rm b}$ & $\log(\rm O3N2)$ $^{\rm b}$  \\
\midrule

$33$ & $4.98_{- 0.93 }^{ +2.28 }$ & $-1.09_{- 0.04 }^{ +0.11 }$& $1.64_{- 0.12 }^{ +0.07 }$\\
$32$ & $7.52_{- 2.07 }^{ +3.58 }$ & $-0.94_{- 0.06 }^{ +0.06 }$& $1.36_{- 0.10 }^{ +0.08 }$\\
$32$ & $15.76_{- 5.85 }^{ +6.67 }$ & $-0.71_{- 0.03 }^{ +0.04 }$& $0.99_{- 0.06 }^{ +0.05 }$\\
$32$ & $38.68_{- 9.23 }^{ +21.78 }$ & $-0.50_{- 0.05 }^{ +0.02 }$& $0.53_{- 0.03 }^{ +0.09 }$\\

\bottomrule
 \end{tabular}
 \end{center}
 \caption{SFR and emission-line ratios of $z\sim1.5$ composite spectra.}
  {$^{\rm a}$}{ Median SFR of galaxies in the bin.}\\
 {$^{\rm b}$}{ Line ratio measured from the composite spectrum.}
 \label{tab:SFRstacks}
\end{table}

\begin{table}
\begin{center}
\renewcommand{\arraystretch}{1.4}
\begin{tabular}{rrr}
\toprule
\multicolumn{3}{c}{$\rm [NII]/H\alpha\ H_{\rm AB} \ stacks$}\\
\midrule
    $N_{\rm gal}$ & $H_{\rm AB,\ med}$ $^{\rm a}$ & $\log(\rm [NII]/H\alpha)$ $^{\rm b}$ \\
\midrule
$30$ & $21.58$ & $-0.56_{- 0.03 }^{ +0.04 }$\\
$30$ & $22.03$ & $-0.62_{- 0.04 }^{ +0.03 }$\\
$30$ & $22.39$ & $-0.63_{- 0.06 }^{ +0.05 }$\\
$30$ & $22.66$ & $-0.72_{- 0.06 }^{ +0.06 }$\\
$30$ & $22.87$ & $-0.83_{- 0.04 }^{ +0.07 }$\\
$30$ & $23.21$ & $-0.94_{- 0.08 }^{ +0.10 }$\\
$29$ & $23.54$ & $-1.06_{- 0.16 }^{ +0.07 }$\\
$29$ & $23.85$ & $-1.19_{- 0.05 }^{ +0.17 }$\\

\toprule
\toprule
\multicolumn{3}{c}{$\rm O3N2\ H_{\rm AB} \ stacks$}\\
\midrule
    $N_{\rm gal}$ & $H_{\rm AB,\ med}$& $\log(\rm O3N2)$  \\
\midrule
$28$ & $21.58$ & $0.51_{- 0.02 }^{ +0.16 }$\\
$28$ & $22.06$ & $0.80_{- 0.07 }^{ +0.11 }$\\
$27$ & $22.41$ & $0.82_{- 0.09 }^{ +0.12 }$\\
$27$ & $22.66$ & $1.08_{- 0.09 }^{ +0.17 }$\\
$27$ & $22.88$ & $1.12_{- 0.08 }^{ +0.13 }$\\
$27$ & $23.21$ & $1.37_{- 0.12 }^{ +0.16 }$\\
$27$ & $23.54$ & $1.54_{- 0.09 }^{ +0.18 }$\\
$27$ & $23.85$ & $1.72_{- 0.17 }^{ +0.11 }$\\

 \bottomrule
 \end{tabular}
 \end{center}
 \caption{$H_{\rm AB}$ and emission-line ratios of $z\sim1.5$ composite spectra.}
 {$^{\rm a}$}{ Median $H_{\rm AB}$ of galaxies in the bin.}\\
 {$^{\rm b}$}{ Line ratio measured from the composite spectrum.}
\label{tab:Hstacks}
\end{table}

\subsection{Implications for Low-Resolution Spectroscopic Surveys}

While strong rest-optical emission line ratios are useful for understanding the gas-phase metallicity and other physical properties of the ISM in high-redshift galaxies, knowledge of the [NII]/H$\alpha$ ratio is required to interpret the results of future large spectroscopic facilities such as the \textit{Roman Space Telescope} and \textit{Euclid}. These observatories will obtain low-resolution ($R\sim435-865$) spectra of a large number of high-redshift galaxies, addressing several outstanding questions in cosmology. Low-resolution spectra will result in a blending of H$\alpha$ with the nearby [NII] lines, presently a problem for future surveys utilizing these facilities.  This blending will bias measurements such as the H$\alpha$ flux or the systemic redshift in an amount that depends on the strength of the [NII] line.  The strength of [NII] relative to H$\alpha$ correlates with other galaxy properties such as stellar mass or SFR, and therefore could bias measurements that also depend on these quantities.  \citet{Faisst2018} have constructed a calibration for the [NII] contamination fraction as a function of galaxy properties resulting from the MZR \citep[see also][ Appendix D]{Reddy2018}.  This calibration presents an improvement over previous works which assumed a constant [NII] contamination across the sample \citep{Colbert2013, Mehta2015}.  However, the data used to construct this calibration \citep{Kashino2017} are not deep enough to probe the full range of [NII]/H$\alpha$ for the $z\sim1.5$ galaxy population.  In addition, \citet{Martens2018} investigates the effects of $\rm [NII]+H\alpha$ contamination on galaxy clustering and cosmological parameters using a $z\sim1.5$ dataset from \citet{Wuyts2016} that is insufficient to constrain the full $z\sim1.5$ galaxy population.

In particular, galaxy clustering measurements derived from galaxy systemic redshifts will be biased based on the blended contribution of [NII] to H$\alpha$, which will shift the measured  blended H$\alpha$ centroid. As we have demonstrated here, the [NII]/H$\alpha$ ratio is strongly correlated with galaxy properties such as stellar mass, SFR, and $H$-band magnitude. The common solution of assuming a constant [NII] contamination fraction of $0.29$ based on average values in the local universe is likely insufficient for describing galaxies at high redshift. The left panel of Figure~\ref{fig:deltas} illustrates the importance of understanding the stellar mass-dependent [NII]/H$\alpha$ contamination based on a sample of $z\sim1.5$ galaxies in that the [NII]/H$\alpha$-$M_*$ relation at high redshift is not only offset compared to in the local universe, but scales differently with stellar mass.  Figures~\ref{fig:NIIothers} and \ref{fig:o3n2others} display N2 and O3N2 line ratios, respectively, as a function of the additional parameters of SFR and $H_{\rm AB}$ magnitude.  As these quantities will be easily measured in future large surveys utilizing low-resolution spectra, they offer appealing alternatives to $\rm M_*$ as the independent measurement for quantifying the effects of blended $\rm [NII]+H\alpha$ emission. The lower number of composite spectra in bins of SFR compared to the $M_*$ stacks reflects the lower sample size of 129 galaxies which requires a $>3 \sigma$ detection in H$\beta$ in order to obtain a robust dust-corrected SFR.  We find that both N2 and O3N2 scale with SFR in a manner that mirrors their scaling with stellar mass, such that increasing SFR corresponds to increasing N2 and decreasing O3N2. This is likely reflecting the positive correlation between SFR and stellar mass.  In addition, we find that N2 decreases (i.e., lower O/H) with increasing (i.e., less luminous) $H_{\rm AB}$ magnitude. Finally, O3N2 behaves such that the O3N2 ratio increases with increasing $H_{\rm AB}$ magnitude. Tables~\ref{tab:SFRstacks} and \ref{tab:Hstacks} give the measurements for our composite spectra in bins of SFR and $H_{\rm AB}$, respectively. In future work, we will utilize our $z\sim1.5$ dataset to understand this blended [NII]/H$\alpha$ contamination as a function of these galaxy properties covering the full range of [NII]/H$\alpha$ ratios spanning the $z\sim1.5$ galaxy population, and estimate the effect of the blending for the determination of cosmological parameters.

\section{Summary and Conclusions}
\label{sec:summary}

In this paper we have analyzed spectra of $z\sim1.5$ galaxies from the MOSDEF survey.  Using N2 and O3N2 line ratios, we investigated the evolution of these line ratios as a function of galaxy properties including the stellar mass, SFR, and $H_{\rm AB}$. We further investigate the secondary dependence  on SFR at fixed stellar mass and evidence of the FMR in our individual galaxies and stacked spectra.  Finally, we discussed two different calibrations used to estimate the oxygen abundance from strong-line ratios, and provide some context for the use of these measurements in next-generation low-resolution wide-field surveys.  A summary of our key results is provided below.

(i) We have demonstrated that the [NII]/H$\alpha$ ratio of galaxies at $z\sim1.5$ with individual detections of these lines, as well as stacked spectra comprising a large sample of galaxies for which only H$\alpha$ is detected, shows a clear correlation with stellar mass. This relation appears to be well fit by Equation~\ref{eqn:equation} which resembles a power law at low mass and flattens out at high masses. In addition, we find that the N2 ratio is correlated with the SFR of galaxies in this sample.

(ii) We find that O3N2 is strongly anti-correlated with stellar mass in our sample of galaxies at $z\sim1.5$. One difference between this line ratio and the results of the N2 relation is that this ratio does not appear to flatten at high masses and is well fit by a single power law. Additionally, this anti-correlation is present as a function of SFR.  

(iii) We established that the deviations from the mean relations of $\log(\rm O/H)_{\rm O3N2}$ and sSFR at fixed stellar mass are anticorrelated with a slope of $-0.09\pm0.06$, indicating the existence of an secondary SFR dependence of the MZR at $z\sim1.5$, though with a limited significance of $1.5\sigma$.  This correlation exists when considering the sample of individual galaxies that have $>3\sigma$ detections in [OIII], H$\beta$, [NII], and H$\alpha$, however including such a restriction may introduce systematics in our sample. This observed anticorrelation is shallower than that observed at $z\sim2.3$ \citep{Sanders2018}, but consistent within $1\sigma$. In addition, this anticorrelation is present when making stacks in $M_*$ and sSFR that include objects without detections in one or more of the lines, suggesting that the evidence for the FMR is real. 

(iv) Using N2 and O3N2 calibrations from \citetalias{PP04} and \citetalias{Bian2018} we find evolution of the MZR between $z\sim2.3$, $z\sim1.5$, and $z=0$.  The MZR at high redshift calculated using N2 does not strongly differ depending on which calibration is used, however the O3N2 MZR shows an offset of $\sim0.1$ dex between the two calibrations.  The evolution of O3N2, and the corresponding O3N2-based metallicity appears to change toward $z=0$ at a constant rate as a function of stellar mass, having an offset of $\Delta\rm log(\rm O/H)_{\rm O3N2}\sim0.30$ dex between $z\sim1.5$ and $z=0$.  In contrast, the difference between the local and $z\sim1.5$ MZR derived from observations of N2 does strongly depend on stellar mass, having an offset of $\Delta\rm log(\rm [NII]/H\alpha)\sim0.45$ dex ($\Delta\rm log(\rm O/H)\sim0.25$ dex) below $M_*\sim 10^{9.75}\rm M_{\odot}$, but only a difference of $\Delta\rm log(\rm [NII]/H\alpha)\sim0.10$ dex ($\Delta\rm log(\rm O/H)\sim0.05$ dex) at $M_*\gtrsim 10^{10.5}\rm M_{\odot}$. The metallicity calibrations that use O3N2 benefit from the large dynamic range of observed line ratios and do not suffer from saturation in star-forming galaxies at high mass. Therefore, the O3N2 calibration if preferred when calculating gas-phase metallicities of our $z\sim1.5$ sample. These different evolutionary results from N2 and O3N2 show that although we observe a secondary dependence of the MZR on SFR, the normalization, and consequently the evolution of the FMR as a function of redshift, depends on how metallicities are calculated. This uncertainty in the normalization of the FMR highlights the importance of using an appropriate metallicity calibration.

(v) Understanding the [NII]/H$\alpha$ ratio of galaxies at $z\sim1.5$ as a function of galaxy properties is very important for future large spectroscopic surveys which will need to disentangle their measurements of a blended [NII]+H$\alpha$ emission lines. These properties, including $\rm M_*$, SFR, and $H_{\rm AB}$, are quantities that will be easily measured in future surveys. Previous studies have corrected for the contamination of [NII] by using either a single value for the entire galaxy population, or a scaling relation based off of local galaxies.  However, we have demonstrated that the N2 relation at $z\sim1.5$ is distinct from that in the local universe, and is not just offset, but has a different shape.  The difference between the N2 ratio at $z\sim1.5$ and that at $z=0$ can be substantial, up to $\sim0.5$ dex depending on the stellar mass.  

The MZR remains a crucial lens through which to view the evolution of galaxies throughout cosmic time. We have investigated how two strong-line ratios evolve from the local universe through $z\sim1.5$ and found significant differences in both N2 and O3N2 compared to at $z=0$.  These differences are critical as a basis for interpreting future results from the \textit{Roman Space Telescope} and \textit{Euclid}, the details of which are forthcoming.  While we have investigated the MZR of our sample at $z\sim1.5$, and presented evidence for the FMR at this epoch, and its consistency with the FMR of local galaxies, the difficulty of measuring accurate metallicities will need to be remedied through a more direct calibration of high-redshift galaxies \citep{Sanders2020}, and applied to samples an order of magnitude larger than the current sample in order to form a more complete galaxy evolution model.

\section*{Acknowledgements}
We acknowledge support from NSF AAG grants AST1312780, 1312547, 1312764, and 1313171, grant AR-13907 from the Space Telescope Science Institute, and grant NNX16AF54G from the NASA ADAP program. We also acknowledge a NASA contract supporting the “WFIRST Extragalactic Potential Observations (EXPO) Science Investigation Team” (15-WFIRST15-0004), administered by GSFC. We thank the 3D-HST collaboration, who provided us with spectroscopic and photometric catalogs used to select MOSDEF targets and derive stellar population parameters. Support for RLS was provided by NASA through the NASA Hubble Fellowship grant \#HST-HF2-51469.001-A awarded by the Space Telescope Science Institute, which is operated by the Association of Universities for Research in Astronomy, Incorporated, under NASA contract NAS5-26555. We wish to extend special thanks to those of Hawaiian ancestry on whose sacred mountain we are privileged to be guests. Without their generous hospitality, most of the observations presented herein would not have been possible.
 
 \section*{Data Availability}
 The data underlying this article are publicly available and can be obtained at \texttt{http://mosdef.astro.berkeley.edu/for-scientists/data-releases/}.

\bibliographystyle{mnras}
\bibliography{mosdef_z1.5_roman}

\end{document}